\documentclass[journal,a4paper]{IEEEtran}

\usepackage{cite}

\usepackage{graphicx}
\usepackage{epstopdf}

\usepackage{pgf}
\usepackage{tikz}
\usepackage[utf8]{inputenc}
\usepackage{pgfplots}
\usetikzlibrary{calc}
\usetikzlibrary{shapes.misc}
\usetikzlibrary{shapes.arrows}
\usetikzlibrary{shapes.symbols}
\usetikzlibrary{positioning}
\usetikzlibrary{arrows.meta}
\usetikzlibrary{shapes.geometric, arrows}
\usepackage{varwidth}
\usepackage{siunitx}

\usetikzlibrary{fit}
\graphicspath{{Images/}}
\DeclareGraphicsExtensions{.pdf,.jpeg,.png,.eps}

\usepackage{amsmath}
\usepackage{amssymb}
\usepackage[short]{optidef}
\usepackage{bm}

\usepackage{siunitx}
\usepackage{algorithmicx}
\usepackage[noend]{algpseudocode}
\usepackage{algorithm}
\usepackage{booktabs}
\algnewcommand{\Initialize}[1]{%
  \State \textbf{Initialize:}
  \Statex \hspace*{\algorithmicindent}\parbox[t]{.8\linewidth}{\raggedright #1}
}

\usepackage{array}

  \usepackage[caption=false,font=footnotesize]{subfig}

\tikzset{
	house/.style={
		draw,
		single arrow,
		single arrow head extend=1,
		minimum height=10,
		shape border rotate=90,
	}
}

\usepackage{stfloats}
\tikzset{
	comp/.style = {
		minimum width  = 8cm,
		minimum height = 4.5cm,
		text width     = 8cm,
		inner sep      = 0pt,
		text           = green,
		align          = center,
		font           = \Huge,
		transform shape,
		thick
	},
	monitor/.style = {draw = none, xscale = 18/16, yscale = 11/9},
	display/.style = {shading = axis, left color = black!60, right color = black},
	ut/.style      = {fill = gray}
}
\tikzset{
	computer/.pic = {
		\node(-m) [comp, pic actions, monitor]
		{\phantom{\parbox{\linewidth}{\tikzpictext}}};
		\node[comp, pic actions, display] {\tikzpictext};
		\begin{scope}[x = (-m.east), y = (-m.north)]
			\path[pic actions, draw = none]
			([yshift=2\pgflinewidth]-0.1,-1) -- (-0.1,-1.3) -- (-1,-1.3) --
			(-1,-2.4) -- (1,-2.4) -- (1,-1.3) -- (0.1,-1.3) --
			([yshift=2\pgflinewidth]0.1,-1);
			\path[ut]
			(-1,-2.4) rectangle (1,-1.3)
			(-0.9,-1.4) -- (-0.7,-2.3) -- (0.7,-2.3) -- (0.9,-1.4) -- cycle;
			\path[pic actions, fill = none]
			(-1,1) -- (-1,-1) -- (-0.1,-1) -- (-0.1,-1.3) -- (-1,-1.3) --
			(-1,-2.4) coordinate(sw)coordinate[pos=0.5] (-b west) --
			(1,-2.4) -- (1,-1.3) coordinate[pos=0.5] (-b east) --
			(0.1,-1.3) -- (0.1,-1) -- (1,-1) -- (1,1) -- cycle;
			\node(-c) [fit = (sw)(-m.north east), inner sep = 0pt] {};
		\end{scope}
	}
}

\tikzset{>=stealth}

\usepackage{url}

\begin{document}
%
\title{Grid-Constrained Distributed Optimization for Frequency Control with Low-Voltage Flexibility}

\author{Jonas~Engels,~\IEEEmembership{Student~Member,~IEEE,}
	Bert~Claessens 
	and~Geert~Deconinck,~\IEEEmembership{Senior~Member,~IEEE}
	
	\thanks{Jonas Engels is with REstore NV/SA, Antwerp, Belgium and with the Department of Electrical Engineering, KU Leuven/EnergyVille, Leuven, Belgium  (jonas.engels@restore.energy)}
	\thanks{Bert Claessens is with REstore, Antwerp, Belgium (bert.claessens@restore.energy)}
	\thanks{Geert Deconinck is with the Department of Electrical Engineering, KU Leuven/EnergyVille,	Leuven, Belgium (geert.deconinck@kuleuven.be)}
	\thanks{This work is partially supported by Flanders Innovation \& Entrepreneurship (VLAIO)}
	\thanks{\copyright 2019 IEEE. Personal use of this material is permitted.  Permission from IEEE must be obtained for all other uses, in any current or future media, including reprinting/republishing this material for advertising or promotional purposes, creating new collective works, for resale or redistribution to servers or lists, or reuse of any copyrighted component of this work in other works.}}


\maketitle

\begin{abstract}
Providing frequency control services with flexible assets connected to the low-voltage distribution grid, amongst which residential battery storage or electrical hot water boilers, can lead to congestion problems and voltage issues in the distribution grid. 
In order to mitigate these problems, a new regulation has been put in place in Belgium, imposing a specific constraint: in any circle with a radius of 100~m, there can be at maximum 10 connection points providing frequency control at any time.

This paper presents an impact analysis and a coordination strategy of a Flexibility Service Provider (FSP) that operates a pool of assets and is exposed to this new regulatory constraint.
Results show that at 5\thinspace\% participation, only 90\thinspace\% of total control capacity can be used, with a large difference between neighbourhoods with different population densities. 

A distributed optimization framework to coordinate the assets arises naturally, in which the assets are able to keep their local cost functions private and only have to communicate with neighbouring assets that are geographically close, and with the FSP. 
Analysis of the proposed distributed optimization algorithm shows a clear trade-off between optimality gap, owing to the mixed-integer nature of the problem, and iterations to convergence.

\end{abstract}

\section{Introduction}\label{sec:intro}
In recent years, ancillary services markets in Europe have been opening up for third party participants and non-conventional energy resources, such as battery storage or industrial demand response. 
The primary frequency control or frequency containment reserve (FCR) market~\cite{ENTSO-E2013}, where one is able to sell power capacity for primary frequency control to the Transmission System Operator (TSO), is one of the markets that have seen an increased participation of these new, flexible energy resources. 
This evolution has also fuelled interest in using flexibility from flexible residential energy resources such as domestic hot water heaters, which are connected to the low-voltage distribution grid.

As these assets are not able to participate in the FCR market by themselves, a new party, the Flexibility Service Provider (FSP) is needed that facilitates the access of these assets to the FCR market, both in a technical and in an administrative way.
The flexible FCR capacity of each of these assets is an order of magnitude smaller than the bid granularity in the FCR markets (e.g. 1 MW in Belgium~\cite{Elia-GF}). Therefore, the FSP will have to put various flexibility assets together in a pool large enough to participate in the FCR market. Thereby, the FSP has to make sure that its pool of assets can provide a constant FCR capacity for the duration of the bid (e.g. one week in Germany on the international Regelleistung bidding platform~\cite{Regelleistung}).

As these new, flexible energy resources are connected to the distribution grid rather than directly to the transmission grid, the distribution system operator (DSO) also becomes a stakeholder. The DSO has to make sure that the distribution grid remains within its operational constraints when these assets perform the FCR service. This is challenging, as the distribution grid has historically not been designed for these kinds of demand response actions.


%

In~\cite{Shahsavari}, it is shown that when some of these assets are located in the same area and are activated synchronously, which is usually the case when providing FCR, this can lead to congestion problems in the low-voltage distribution grid.
Congestion in the distribution grid occurs when the transfer of active power over the grid exceeds the transfer capability of the grid, which is limited by the operational grid constraints: voltage limits, thermal limits of cables and transformers, the interface with the TSO and protection equipment \cite{Andersen2012}.

Performing a detailed grid study on the impact of using each of these assets for FCR is too time-consuming, costly and requires detailed grid information, which is often not available.
Therefore, various solutions to distribution grid congestion have been proposed in the literature, such as voltage regulation with active and reactive power control \cite{TantGeth_2013, d2010distributed, Demirok_2011}.
A local voltage droop controller is presented in~\cite{DECONINCK2015}, which is shown to be effective in avoiding distribution grid constraint violations while having very limited impact on the performance of the service to be delivered. 

However, straightforward power curtailment cannot be applied to the FCR service, as this would result in non-delivery of the service to the TSO and hence into penalties for the FSP. 
Controlling the reactive power output of the grid-connected inverters could also reduce voltage issues~\cite{Deshmukh2012}. However, this results in increased resistive losses as injecting additional reactive power increases the current through the cables~\cite{Demirok_2011}. An optimal control minimizing these losses is rather complex and requires additional control logic to be installed~\cite{Deshmukh2012}.

As these methods have their drawbacks, the DSO is looking at new regulations that are easily enforceable to avoid distribution grid problems with assets providing FCR.




The remainder of the paper is organized as follows: Section~\ref{sec:FCR_BE} explains the new 2018 Belgian regulation on providing FCR with low-voltage grid connected assets and motivates the distributed optimization architecture proposed in this paper. Section~\ref{sec:circles} describes an algorithm to construct the relevant constraints imposed by these new regulation. Section~\ref{sec:distr_opt} then formulates the central optimization problem and the distributed optimization algorithm for an FSP that is exposed to this regulation. Section~\ref{sec:casestudy} evaluates the impact of the new regulation and the performance of the distributed optimization. Finally, the paper is concluded in Section~\ref{sec:conclusion}.

\section{FCR with Low-Voltage Connected Assets in Belgium}\label{sec:FCR_BE}
Recently, the Belgian federation of electrical and gas network operators, Synergrid~\cite{synergrid}, has proposed a standard agreement contract between DSOs and FSPs that want to exploit flexibility on the low-voltage distribution grid for FCR services~\cite{synergrid_agreement}.
The agreement presents some constraints by which the FSP should comply in order to prevent congestion issues in the distribution grid when using the flexibility for FCR. The proposed constraints in the document are easily enforceable and do not require complicated assessments such as a detailed power flow calculation. 

The two constraints imposed by Synergrid in the agreement contract 
are the following:

\begin{enumerate}
	\item The maximum flexible power capacity used for FCR at one low-voltage connection point is \SI{5}{kW}.
	\item Within any circle with a radius of \SI{100}{\metre}, there can be a maximum of 10 low-voltage connection points in the pool of the FSP  providing flexibility for FCR at the same time.
\end{enumerate}
The first constraint is straightforward and does not require further explanation. The second constraint is slightly more complicated and creates some room for optimization by the FSP. If, for instance, the FSP has 20 assets in its pool that are all located within a circle with radius of \SI{100}{\metre}, the FSP can choose which of the 20 assets should provide the FCR capacity at each moment in time. 
It would then be beneficial to select the assets that can provide the \emph{cheapest} FCR capacity at each moment in time. 
Besides, when assets are located in multiple circles it is not straightforward to select which assets should deliver FCR at minimal costs, as each circle imposes its constraint and all of them should be respected.

To find the cheapest FCR capacity, one has to define the cost of providing FCR capacity with a flexibility asset. 
This cost can include both the actual marginal cost of providing the flexibility (e.g. degradation cost of a battery providing frequency control) and the opportunity costs of using the same flexibility for other purposes (e.g. using the battery to store locally generated PV energy).
Optimizing in this way can lead to increased revenues for all parties, as 
synergies exist by combining flexibility for different objectives such as frequency control and electricity tariff optimization~\cite{LITJENS2018172,Engels2017}.

\subsection{A Distributed Optimization Framework for the FSP}\label{sec:dist}
As the flexible capacity from one asset connected to the distribution grid is usually rather small (and explicitly limited to \SI{5}{kW} by the first constraint of Synergrid), there need to be a large number of assets in the pool of an FSP. 
This also means that, in case the entire optimization is performed centrally by the FSP, it can quickly become intractable due to the high number of variables and constraints~\cite{Chapman2016}. 

A well-studied approach to mitigate this intractability is to distribute the optimization problem amongst the various assets in the FSP pool. This has the advantage that each asset only has to solve a small, local optimization problem. Besides, the assets can implement their constraints and cost functions locally, keeping this information private from the other assets and from the FSP, which is favourable from a confidentiality point of view~\cite{Mhanna_2016}.
Finally, a distributed optimization architecture arises naturally here, as the second constraint imposed by Synergrid, limiting the number of active assets in each circle of \SI{100}{\metre}, results in a multitude of constraints, each including only neighbouring assets which are geographically close together.

In the literature, various architectures of distributed demand response aggregation have been proposed. A non-iterative, distributed approach is presented in~\cite{Weckx_MASchargin}, in which the assets calculate
their local costs in a distributed way for each possible value of the dual variables.
However, this works only in case the problem is completely decomposable in time, which is not the case here.
Dual decomposition is used in~\cite{Gatsis2013} to aggregate demand response resources while maintaining user confidentiality.
The alternating direction method of multipliers (ADMM), comparable to the distributed method proposed in this paper, is used in~\cite{Wen2012} to optimize electrical vehicle charging while taking into account maximum power constraints of the grid.

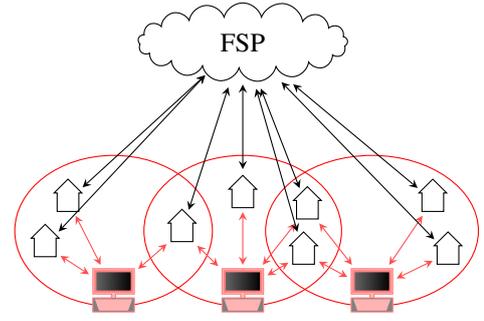
\begin{figure}
	\centering
	\begin{tikzpicture}
	\node [house,scale=1.15, outer sep=0.05cm](a) at (-0.5,0.3) {};
	\node [house,scale=1.15,outer sep=0.05cm](b) at (1.3,0.5) {};
	\node [house,scale=1.15,outer sep=0.05cm](c) at (-0.2,0.9) {};
	
	\node [house,scale=1.15,outer sep=0.05cm](f) at (2.9,0.2) {};
	\node [house,scale=1.15,outer sep=0.05cm](h) at (4.8,0.2) {};
	\node [house,scale=1.15,outer sep=0.05cm](i) at (2.95,0.8) {};
	\node [house,scale=1.15,outer sep=0.05cm](j) at (4.6,0.9) {};
	
	\node [house,scale=1.15,outer sep=0.05cm](k) at (2.1,0.95) {};
		
	\draw[red] (0.4,0.5) ellipse (1.3cm and 1cm);
	\draw[red] (3.8,0.5) ellipse (1.4cm and 1cm);
	\draw[red] (2.1,0.5) ellipse (1.3cm and 1.0cm);

	\pic(comp0) [draw = red!50, fill = red!30, scale=0.06,outer sep=0.05cm] at 	(0.4,-0.17)	{computer};
	\pic(comp1) [draw = red!50, fill = red!30, scale=0.06,outer sep=0.05cm] at 	(3.8,-0.17)	{computer};
	\pic(comp2) [draw = red!50, fill = red!30, scale=0.06,outer sep=0.05cm] at 	(2.1,-0.17)	{computer};

	\node [cloud, cloud puffs=17.2, aspect=4, draw,outer sep=0.1cm](fsp) at (2.1,3) {FSP};
	
	\draw[<->, red!70]   (a) -- (comp0-c);
	\draw[<->, red!70]   (b) -- (comp0-c);
	\draw[<->, red!70]   (c) -- (comp0-c);
	
	\draw[<->, red!70]   (f) -- (comp1-c);
	\draw[<->, red!70]   (h) -- (comp1-c);
	\draw[<->, red!70]   (i) -- (comp1-c);
	\draw[<->, red!70]   (j) -- (comp1-c);
	
	\draw[<->, red!70]   (b) -- (comp2-c);
	\draw[<->, red!70]   (f) -- (comp2-c);
	\draw[<->, red!70]   (i) -- (comp2-c);
	\draw[<->, red!70]   (k) -- (comp2-c);
	
	\draw[<->]   (a) -- (fsp);
	\draw[<->]   (b) -- (fsp);
	\draw[<->]   (c) -- (fsp);
	
	\draw[<->]   (f) -- (fsp);
	\draw[<->]   (h) -- (fsp);
	\draw[<->]   (i) -- (fsp);
	\draw[<->]   (j) -- (fsp);
	\draw[<->]   (k) -- (fsp);
	\end{tikzpicture}
	\caption{Proposed distributed optimization architecture. Each house represent a local flexibility asset and is constraint by circles with a radius of \SI{100}{\metre}, here drawn in red. The assets communicate with the FSP and with circle constraint agents, represented by the red computers.}
	\label{fig:dist_arg}
		\vspace*{-0.3cm}
\end{figure}

Figure~\ref{fig:dist_arg} shows the distributed optimization architecture proposed in this paper, tailored to the problem of the FSP providing FCR while respecting the Synergrid constraints. 
Each asset performs a local optimization, maximizing their FCR revenues while taking into account their local costs and constraints. 
The assets share \emph{circle constraints} with neighbouring assets that are within \SI{200}{\metre} distance, corresponding to the circles in Synergrid's second constraint and illustrated by the red circles in the figure. 

As the assets do not know nor can control the FCR capacity of the neighbouring assets, they cannot enforce the Synergrid's second constraint in the local optimization. 
Therefore, each circle constraint is assigned a \emph{circle constraint agent}, illustrated by the red computers in Figure~\ref{fig:dist_arg}, that ensures there are not more than 10 assets delivering FCR at the same inside the particular circle.
As one asset can be constrained by multiple circle constraints and each circle is managed by only one circle constraint agent, an asset communicates with all circle constraint agents of all circles it belongs to. In this way, it is ensured the asset respects all circle constraints they belong to.
The assets also communicate with the FSP, which coordinates the assets to make sure the sum of the local FCR capacities is constant over the duration of the bidding period. 

With this architecture, no entity has a global view on the central optimization problem, which is distributed amongst all relevant entities, each solving only a small, local part of the problem.

A peer-to-peer architecture, such as presented in~\cite{Engels2016}, 
can also be achieved with the distributed optimization algorithm proposed in this paper. In such a peer-to-peer architecture, each asset would have a local implementation of all circle constraint agents of the circles constraining the asset.
This eliminates the need for circle constraint agents as distinct entities, as each agent would already be implemented locally in the assets constrained by the respective circle. 
Instead of communicating with the circle constraint agents, an asset will then only have to communicate with neighbouring assets with which it shares a 
circle constraint, which are at maximum \SI{200}{m} away.


Transforming the communication with the FSP into a peer-to-peer architecture, thereby eliminating the FSP as a singly point of failure, is a bit more challenging, but can be achieved following the approaches presented in \cite{Engels2016, Mota}. 
In this case, copies of the calculation performed by the FSP have to be implemented locally in some or in all nodes. These nodes can than take the role of the FSP, coordinating the assets towards a constant FCR capacity for the duration of the bid.
To avoid communication between all nodes (all-to-all communication) at every iteration, peer-to-peer communication can be achieved with a gossiping algorithm~\cite{Engels2016} or the D-ADMM~\cite{Mota}. 

The distributed or the peer-to-peer architecture fits perfectly with recently proposed device-to-device communication architectures~\cite{GANDOTRA20179} and the Internet of Things (IoT) paradigm~\cite{IoT}. 
Low Power Wide Area Networks (LPWAN)~\cite{LPWAN} seem to be ideal candidates for this type of communication, as they have low hardware cost, low power consumption and a range largely surpassing the required \SI{200}{\metre}.

The main contributions of this paper can be summarized as follows:
\begin{itemize}
	\item We propose an algorithm to determine all relevant circles according to the new regulatory requirement from Synergrid, which is, to the best of our knowledge, the first time distribution grid constraints are explicitly imposed on demand response flexibility.
	\item We analyse the impact of these constraints on the total amount of FCR capacity that can be offered with a pool of assets connected to the distribution grid, using real data from a DSO.
	\item We describe the mixed-integer optimization problem of an FSP operating a pool of low-voltage grid connected assets providing FCR and present the use of a distributed optimization to solve the problem in a scalable way while keeping costs and constraints of the participating assets confidential.
\end{itemize}



\section{Construction of Circle Constraints}\label{sec:circles}
To be able to implement the optimization problem of the FSP as a mathematical program, we have to translate Synergrid's second constraint into a closed mathematical expression. 
Therefore, we have to be able to find all circles with a radius of \SI{100}{m} that contain at least one connection point. Below, we explain how we can find these circles.
In what follows, we denote a vector by a bold symbol: $\bm{x} = (x_1,x_2, \ldots,x_n)^\intercal$.

We assume the geographical location of all assets or points $\mathcal{I} = \{1,\ldots, n_\mathcal{I}\}$ in the FSP pool is given by their two-dimensional coordinates $\bm{x}_i = (x_i^0, x_i^1), \: i \in \mathcal{I}$ in a two-dimensional Cartesian coordinate system, such that a vector $\bm{x}_i : \Vert\bm{x}_i\Vert^2_2=1$ has a length of 1 meter.
In practice, both the DSO and the FSP should know the geographical location of the participating assets, as the connection points are part of the distribution grid, and the FSP should have a bilateral contract with the owner of the asset allowing the FSP to use the asset for FCR services.

The goal is to find all sets of points $\mathcal{C}_s \subset \mathcal{I}$, of which the smallest circle containing all points in the set has a radius $r$ smaller than or equal to \SI{100}{\metre}, and which is not a subset of any other such set of points:
\begin{IEEEeqnarray}{rCl}\label{eq:circle_set}
\mathcal{C}_s &=& \{i\in \mathcal{I} \mid \exists \bm{c} \in \mathbb{R}^2: \forall i,  \Vert \bm{x}_i-\bm{c} \Vert_2 \leq 100 \} \\
& \text{and}&  \quad  \mathcal{C}_s\nsubseteq \mathcal{C}_{s'}, s'=\{1,\ldots 
,s-1, s+1,\ldots n_\mathcal{S}\}, \IEEEnonumber
\end{IEEEeqnarray}
with $n_\mathcal{S}$ the total number of sets $\mathcal{C}_s$.
The last requirement avoids adding trivial sets of points: for instance, if there is a circle with radius $r\leq\SI{100}{\metre}$ containing points $\{1,2,3\}$, then there are also circles with radii $r\leq\SI{100}{\metre}$ containing only points $\{1,2\}$, $\{2,3\}$ and $\{1,3\}$. However, the constraints that would be imposed by these last three sets of points are already incorporated by the constraint defined by the set $\{1,2,3\}$. Hence, the smaller sets can be discarded.

A naive construction methodology for $\mathcal{C}_s$ would be to check the 
smallest circumscribed circle of all possible combinations of points. However, 
this would quickly become intractable for a rather limited number of points $n_\mathcal{I}$, as the number of possible combinations increases 
exponentially with $O(2^{n_\mathcal{I}})$.

\begin{figure}
	\centering
	
	\begin{tikzpicture}
	
	\def \xix{1.2}
	\def \xiy{0.0}
	\def \xjx{0.7}
	\def \xjy{2.5}
	\def \r{2.2}
	\def \dotwidth{4pt}
	\def \mirrorlinespace{0.3}
	
	\pgfmathsetmacro{\norm}{sqrt((\xix-\xjx)^2+(\xiy - \xjy)^2)}
	\pgfmathsetmacro{\dij}{sqrt(\r^2 - (\norm/2)^2)}
	
	\pgfmathsetmacro{\deltaijx}{(\xix - \xjx)/\norm}
	\pgfmathsetmacro{\deltaijy}{-(\xiy-\xjy)/\norm}
	
	\pgfmathsetmacro{\cix}{(\xix+\xjx)/2+\dij*\deltaijy}
	\pgfmathsetmacro{\ciy}{(\xiy+\xjy)/2+\dij*\deltaijx}
	\pgfmathsetmacro{\cjx}{(\xix+\xjx)/2-\dij*\deltaijy}
	\pgfmathsetmacro{\cjy}{(\xiy+\xjy)/2-\dij*\deltaijx}
	
	\pgfmathsetmacro{\lix}{(\xix+\xjx)/2+(\r+\mirrorlinespace)*\dij*\deltaijy}
	\pgfmathsetmacro{\liy}{(\xiy+\xjy)/2+(\r+\mirrorlinespace)*\dij*\deltaijx}
	\pgfmathsetmacro{\ljx}{(\xix+\xjx)/2-(\r+\mirrorlinespace)*\dij*\deltaijy}
	\pgfmathsetmacro{\ljy}{(\xiy+\xjy)/2-(\r+\mirrorlinespace)*\dij*\deltaijx}
	
	\pgfmathsetmacro{\mx}{(\xix+\xjx)/2}
	\pgfmathsetmacro{\my}{(\xiy+\xjy)/2}

	\draw (\cix, \ciy) circle (\r);
	
	\draw (\cjx, \cjy) circle (\r);
	
	\draw [dashed] (\ljx, \ljy) -- (\lix, \liy) node [pos=0.17, above, sloped] (TextNode) {\textit{mirror line}};
	
	\draw [densely dotted, line width=0.7pt, shorten <=-1cm, shorten >=-1cm] (\xix, \xiy) -- (\xjx, \xjy);
	
	\draw [->, red, line width=1.5pt] (\mx, \my) -- (\mx+\deltaijy, \my+\deltaijx) node [pos=0.9, above=-2pt, sloped] {$\left(\Delta_{ij}^1,\Delta_{ij}^0 \right) $};

	\node(x0) [circle, fill=black,inner sep=0pt,minimum size=5pt, label=left:{$\bm{x}_i$}] at (\xix,\xiy) {};
	
	\node(x1) [circle, fill=black,inner sep=0pt,minimum size=5pt, label=right:{$\bm{x}_j$}] at (\xjx, \xjy) {};
	
	\draw [<->, line width= 1pt, shorten <= 2.5pt, shorten >= 2.5pt] (\mx, \my) -- (\cjx, \cjy) node [midway, above=-2pt, sloped, label distance=-4mm] {$d_{ij}$};
	
	\draw [<->, shorten <= 2.5pt, shorten >= 2.5pt] (\cix, \ciy) -- ($(\cix, \ciy) + cos(50)*\r*(1,0)  + sin(50)*\r*(0,1)$) node [midway, above=-2pt, sloped] {$r$};
	
	\draw [<->, shorten <= 2.5pt, shorten >= 2.5pt] (\cjx, \cjy) -- ($(\cjx, \cjy) + cos(140)*\r*(1,0)  + sin(140)*\r*(0,1)$) node [midway, above=-2pt, sloped] {$r$};
	
	\node(c1) [circle, fill=black,inner sep=0pt,minimum size=\dotwidth, label={[label distance=-2mm]below right:{$\bm{c}_{ij}^1$}}] at (\cix, \ciy) {};
	\node(c2) [circle, fill=black,inner sep=0pt,minimum size=\dotwidth, label={[label distance=-1pt]below:{$\bm{c}_{ij}^2$}}] at (\cjx, \cjy) {};
	
	\node(m) [circle, fill=black,inner sep=0pt,minimum size=\dotwidth, label={[label distance=-1mm]below right:{$\bm{m}_{ij}$}}] at (\mx, \my) {};
	
	\end{tikzpicture}
	
	\caption{Illustration of equations (\ref{eq:circle_middelpoint}) for the construction of the two unique circles with radius $r$ passing through points $\bm{x}_i$ and $\bm{x}_j$.}
	\label{fig:circles_overview}
	\vspace*{-0.3cm}
\end{figure}
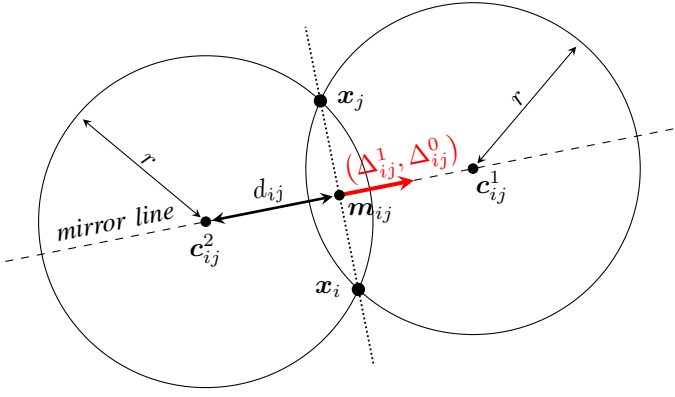

To overcome this, we have developed Algorithm \ref{alg:circle}, which has complexity $O(n_\mathcal{I}^4)$. 
The algorithm is based on the idea that the smallest circumscribed circle of a  set of points has at least two points on the boundary of that circle~\cite{jung1910}. Thus, by finding all circles with radius \SI{100}{\metre} that have at least two points on their boundary, one obtains all circles to be considered when creating the sets $\mathcal{C}_s$.

Given a radius $r$ and two distinct points $\bm{x}_i, \bm{x}_j$, one can define 
two unique circles with centre points $\bm{c}_{ij}^1, \bm{c}_{ij}^2$ as illustrated by Figure \ref{fig:circles_overview}. These centre points can be found using the following equations, resulting from the relations defined in Figure \ref{fig:circles_overview},:

\begin{IEEEeqnarray}{rCl}
\IEEEyesnumber\label{eq:circle_middelpoint} \IEEEyessubnumber*
d_{ij} &=& \sqrt{r^2-\left(  \frac{\Vert \bm{x}_i-\bm{x}_j \Vert_2}{2}\right)^2}, \\
\Delta_{ij}^0 &=&  \frac{x_i^0-x_j^0}{\Vert \bm{x}_i-\bm{x}_j \Vert_2}, \quad  \Delta_{ij}^1 =  \frac{-(x_i^1-x_j^1)}{\Vert \bm{x}_i-\bm{x}_j \Vert_2}, \\
\bm{c}^1_{ij} &=&\left(\frac{x_i^0+x_j^0}{2} + d_{ij} \Delta_{ij}^1, \frac{x_i^1+x_j^1}{2} + d_{ij} \Delta_{ij}^0 \right) ,  \label{eq:centrepoint1}\\  	
\bm{c}^2_{ij} &=&\left(\frac{x_i^0+x_j^0}{2} - d_{ij} \Delta_{ij}^1, \frac{x_i^1+x_j^1}{2} - d_{ij} \Delta_{ij}^0 \right). \label{eq:centrepoint2}
\end{IEEEeqnarray}
Here, $d_{ij}$ gives the distance between the centre points and the midpoint $m_{ij}$ between $\bm{x}_i, \bm{x}_j$ along the \emph{mirror line}, the line with all points at equal distance from both points $\bm{x}_i, \bm{x}_j$. The direction of the mirror line is given by the normalized vector $\left(\Delta_{ij}^1,\Delta_{ij}^0 \right)$. 
The centre points are then found by starting from the midpoint $\bm{m}_{ij} = \left((x_i^0+x_j^0)/2, (x_i^1+x_j^1)/2 \right)$ between $\bm{x}_i, \bm{x}_j$ and going with distance $d_{ij}$ along the mirror line in the positive and the negative direction, as elaborated in equations (\ref{eq:centrepoint1}) and (\ref{eq:centrepoint2}).

\begin{algorithm}
	\caption{Construction of sets within circles of $r\leq\SI{100}{\metre}$}
	\label{alg:circle}
	\begin{algorithmic}[1]
		\ForAll{point $i \in \mathcal{I} $,} \label{alg_s:iter_i}  (in parallel)
		\State $\mathcal{S}_i \gets \emptyset$
		\State $\mathcal{I}_i \gets \{j \in \mathcal{I}  \mid \Vert \bm{x}_i-\bm{x}_j \Vert_2 \leq 200 \}$ \label{alg_s:iter_j}
		\If{$\mathcal{I}_i = \emptyset$}
		\State $\mathcal{S}_i \gets \{i\}$
		\Else  
		\ForAll {$j \in \mathcal{I}_i $,}
		\State Calculate $c^1_{ij}, c^2_{ij}$ from $\bm{x}_i$ and $\bm{x}_j$ using (\ref{eq:circle_middelpoint}).  \label{alg_s:get_circle}
		\State $\mathcal{C}^1 \gets \{n\in \mathcal{I}_i \mid \Vert  \bm{x}_n - \bm{c}^1_{ij} \Vert \leq 100 \}$  \label{alg_s:circle1}
		\State $\mathcal{C}^2  \gets \{n\in \mathcal{I}_i \mid \Vert  \bm{x}_n - \bm{c}^2_{ij} \Vert \leq 100 \}$ \label{alg_s:circle2}
		\ForAll{$\mathcal{C}_s \in \mathcal{S}_i$,} 
		\If{$\mathcal{C}_s \subset \mathcal{C}^1$ or $\mathcal{C}_s \subset \mathcal{C}^2$ }
		\State $\mathcal{S}_i \gets \mathcal{S}_i \setminus \mathcal{C}_s$
		\EndIf
		\EndFor
		\If{$\mathcal{C}^1 \nsubseteq \mathcal{C}_{s},  \forall  \mathcal{C}_{s} \in \mathcal{S}_i$}
		\State $\mathcal{S}_i \gets \mathcal{S}_i \cup \mathcal{C}^1$
		\EndIf			
		\If{$\mathcal{C}^2 \nsubseteq \mathcal{C}_{s},  \forall  \mathcal{C}_{s} \in \mathcal{S}_i$}
		\State $\mathcal{S}_i \gets \mathcal{S}_i \cup \mathcal{C}^2$
		\EndIf			
		\EndFor
		\EndIf
		\EndFor
		\State $\mathcal{S} = \bigcup\limits_{i\in \mathcal{I}} \mathcal{S}_i $ \label{alg_s:union}
	\end{algorithmic}
\end{algorithm}

Algorithm \ref{alg:circle} shows how to construct the set $\mathcal{S} = \{\mathcal{C}_1,\ldots,  \mathcal{C}_{n_\mathcal{S}} \}$ containing all sets $\mathcal{C}_s$ defined by (\ref{eq:circle_set}). The iteration over every asset in step \ref{alg_s:iter_i} creates the local set $\mathcal{S}_i = \{\mathcal{C}_s \in \mathcal{S} | i \in \mathcal{C}_s\}$ containing the circle sets $\mathcal{C}_s$ in which asset $i$ is contained. By executing this iteration in parallel at every asset $i$, the algorithm can be executed in a fully distributed fashion.

The iterations in step \ref{alg_s:iter_j} finds then all neighbouring points $ j$ that are less than or equal to \SI{200}{\metre} apart from each other, as points that are farther from each other can never be in the same circle with radius \SI{100}{\metre}. This limits the combinations to be considered at each point $i$ to the points that are in the neighbourhood of $i$, speeding up up the algorithm significantly. 

Step \ref{alg_s:get_circle} calculates the centre points of the two circles with radius \SI{100}{\metre} determined by points $i, j$. Then, steps \ref{alg_s:circle1} and \ref{alg_s:circle2} determine all points from $\mathcal{I}$ that are enclosed by these circles. This gives two potential sets of points $\mathcal{C}^1, \mathcal{C}^2$, for which it has to be checked if there does not already exist a set $\mathcal{C}_s \in \mathcal{S}_i$ that is a subset of  $\mathcal{C}^1$ or $\mathcal{C}^2$, in which case $\mathcal{C}_s$ is removed from $\mathcal{S}_i$. Finally, if the sets $\mathcal{C}^1, \mathcal{C}^2$ are not in itself a subset of any $\mathcal{C}_s \in \mathcal{S}_i$, they are added to $\mathcal{S}_i$. These last two operations are performed to eliminate trivial sets, explained above.
Finally, In step \ref{alg_s:union}, the set $\mathcal{S} = \{\mathcal{C}_1, \ldots, \mathcal{C}_{n_\mathcal{S}}\}$ is created by taking the union over all subsets $\mathcal{S}_i$. However, when using the distributed optimization algorithm presented further this step is not required as in the local optimization problem (\ref{eq:local_opt}) each asset $i$ only needs the information of the subset $\mathcal{S}_i$.

\begin{figure}
	\centering
	\includegraphics[scale=0.95]{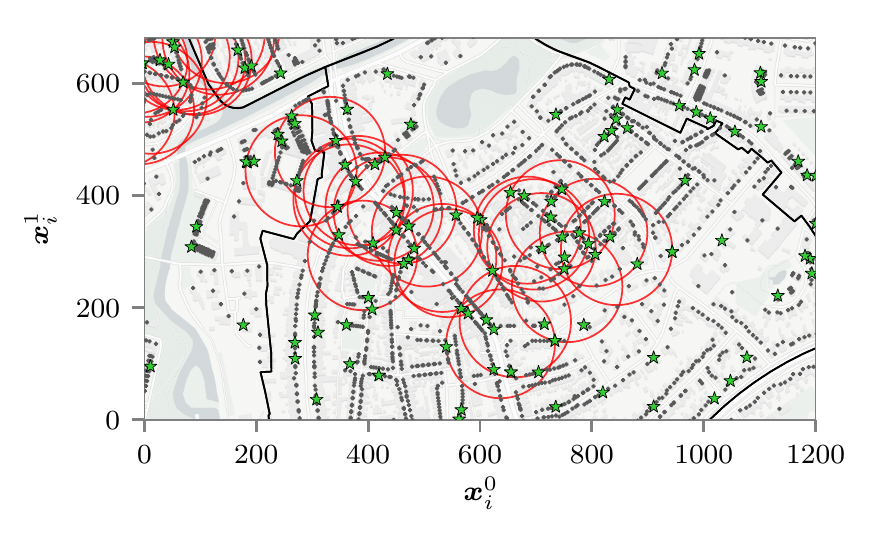}
	\vspace*{-0.5cm}
	\caption{Example of the circles with more than 10 assets ($|\mathcal{C}_s| > 10$) in the \emph{Zandberg} neighbourhood in the city of Breda. The black dots are all the connection points in the neighbourhood, obtained from \cite{Enexis_data}, while the green stars denote the randomly selected connection points participating in the pool of the FSP, corresponding to \SI{5}{\%} of all connection points. Map data \copyright~OpenStreetMap contributors}
	\vspace*{-0.3cm}
	\label{fig:circles}
\end{figure}

An example of the results of the algorithm, applied to a neighbourhood in the city of Breda, is given in Figure~\ref{fig:circles}. As one can see, the closer the points are together, the more circles can be drawn and thus more constraints have to be applied. 

\section{Distributed Optimization of a Pool of Assets}\label{sec:distr_opt}
This section first formulates the centralized optimization to be performed by an FSP that wants to sell FCR capacity to the TSO over the duration of one bidding period $n_\mathcal{T}$ with a pool of flexible assets connected to the distribution grid.
Subsequently, this section elaborates the proposed distributed optimization algorithm, solving the problem according to the distributed architecture presented in Figure~\ref{fig:dist_arg}.

\subsection{Central Optimization Problem}
The objective of the optimization problem to be performed by the FSP is to maximize revenues from selling FCR capacity $p_{F}$ provided by a pool of assets, minus the costs of using these asset for primary frequency control.

In European FCR markets, 
the FSP only gets paid a capacity fee (and no activation fee) when providing FCR to the TSO~\cite{Regelleistung}. 
Therefore, the revenues for providing FCR are given by $c_{F} p_{F} n_\mathcal{T}$, with $c_{F}$ the unit price to provide FCR per time step $t\in \mathcal{T} = \{1,\ldots,n_\mathcal{T}\}$, $n_{\mathcal{T}}$ the number of time steps in the bidding period, and $p_{F}$ the aggregated FCR capacity the FSP is able to sell to the TSO. 
As the FCR price $c_{F}$ and the amount of FCR capacity sold on the market $p_{F}$ should be constant over the entire bidding period~\cite{Elia-GF}, $c_{F}$ and $p_{F}$ do not depend on the time~$t$.

We define $\bm{p}_i  = (p_{i,0},\ldots, p_{i,n_\mathcal{T}})^\intercal$ as the vector containing the FCR capacities in kW provided by asset $i$ at every time step $t$ of the bidding period.
As the FSP sells the aggregate of these local FCR capacities to the TSO, the sum of the local FCR capacities over all assets should equal to the total FCR capacity sold $p_F$ for every time step: $\sum_{i\in\mathcal{I}} p_{i,t}= p_F,  \forall t \in \mathcal{T}$.

We define also $c_i : \mathbb{R}^{n_{\mathcal{T}}} \mapsto \mathbb{R} \cup \{+ \infty\}$ as the local cost $c_{i}(\bm{p}_{i})$ of asset $i$ to provide the FCR capacity vector $\bm{p}_{i}$.
As explained in Section II, this cost function can include both the actual marginal cost of controlling the asset 
and the opportunity costs of using the flexibility for other purposes. 
Such a cost function can also be viewed as the negative of a utility function, used in previous work~\cite{Samadi2012, Weckx_MASchargin}.
We allow $c_i(\bm{p}_{i})$ to take on the value $+ \infty$ when the point $\bm{p}_i$ is infeasible for the asset (e.g. a \SI{2}{kW} battery cannot provide \SI{5}{kW} frequency control capacity). 
To ensure a global optimum can be found, we assume $c_i(\bm{p}_{i})$ to be convex.

The complete optimization problem to be solved by the FSP can then be formulated as a mixed-integer optimization program:
\begin{mini!}
{\bm{p}_i, \bm{z}_i, p_{F}}{ \sum_{i\in\mathcal{I}}  c_{i}(\bm{p}_{i})  -   c_{F} p_{F} n_{\mathcal{T}} \label{eq:opt_problem_obj}}
{\label{eq:opt_problem}}{}
\addConstraint{0\leq {p}_{i,t} }{  \leq 5 z_{i,t} , \quad}{\forall t \in \mathcal{T} ,  \forall i \in \mathcal{I}  \label{eq:opt_problem_constr}}
\addConstraint{\sum_{i\in\mathcal{I}} p_{i,t}}{= p_F,  }{ \forall t \in \mathcal{T} \label{eq:opt_constr_global}}
\addConstraint{\sum_{i\in\mathcal{C}_s} z_{i,t}}{\leq 10, \quad}{\forall t \in \mathcal{T} ,  \forall \mathcal{C}_s \in \mathcal{S} \label{eq:opt_constr_circle}}
\addConstraint{\bm{z}_i }{ \in \{0, 1\}^{n_\mathcal{T}}, \quad  }{\forall i \in \mathcal{I}. \label{eq:opt_constr_bool}}
\end{mini!}
Here $\bm{z}_i  = (z_{i,0},\ldots, z_{i,n_\mathcal{T}})^\intercal$ is a vector of binary variables $z_{i,t}$ which gives $1$ if asset $i$ is providing FCR capacity at time step~$t$ and $0$ otherwise. 
Constraint~(\ref{eq:opt_problem_constr}) represents the first constraint of Synergrid, limiting the FCR capacity to \SI{5}{kW} in case the asset is delivering FCR capacity (i.e.  $z_{i,t} = 1$).
Constraint~(\ref{eq:opt_constr_global}) represents the requirement that the sum of the local FCR capacities should equal the total FCR capacity the FSP sells to the TSO, at each time step.
Constraint~(\ref{eq:opt_constr_circle}) represents the requirement to have maximum 10 assets providing FCR capacity in each circle of \SI{100}{\metre}. Finally, (\ref{eq:opt_constr_bool}) constraints $\bm{z}_{i}$ to a binary variable of dimension $n_{\mathcal{T}}$.

Problem~(\ref{eq:opt_problem}) is a mixed-integer optimization with a convex continuous relaxation, for which various solution methods exists that are able to find the global optimum, e.g. branch-and-bound~\cite{BranchandBound1985}, the extended cutting plane method~\cite{WESTERLUND1995131} or the branch-and-cut method~\cite{Stubbs1999}.

However, as this problem contains $n_\mathcal{T} n_\mathcal{I}$ binary variables, the complexity increases quickly with a growing number of assets. 
Therefore, we create a distributed version of the optimization problem~(\ref{eq:opt_problem}), in which the assets only have to communicate with the FSP and 
the applicable circle constraint agents (or with their local neighbours in a peer-to-peer architecture, as explained in Section \ref{sec:dist}).

When participating in the FCR market, the FSP has to bid in the FCR capacity $p_F$ at a certain capacity price $c_F$. 
The TSO then selects the cheapest bids in merit order, until the required FCR capacity is reached. 
As the market is a pay-as-bid market, the FSP only gets paid his bid price $c_F$ and not the clearing price~\cite{Regelleistung}. 
The FSP will thus first have to decide on the price $c_{F}$, which should be high enough to obtain as much revenues as possible, but not too high as then the bid might not be accepted. 
With the bid price $c_{F}$ decided, the FSP can use (\ref{eq:opt_problem}) to optimize the FCR capacity of its pool of assets.
%

\subsection{Distributed Optimization}


One can identify three parts in problem (\ref{eq:opt_problem}): a local optimization to maximize the local FCR revenues minus the local costs, the global problem of the FSP, who tries to obtain a constant FCR capacity from all assets over the bidding period $n_{\mathcal{T}}$, and the local circle constraints imposed by Synergrid.

These three parts give a natural way to distribute the optimization problem into three subproblems. The first subproblem is the local optimization per asset to maximize $f_i(\bm{p}_i^f, \bm{z_i}^f)$, the local FCR revenues minus the local cost, equal to the objective (\ref{eq:opt_problem_obj}) constrained to (\ref{eq:opt_problem_constr}).
The second subproblem is the optimization performed by the FSP, minimizing $h(\bm{p}^h_{i\in\mathcal{I}})$, the indicator function corresponding to (\ref{eq:opt_constr_global}):
\begin{equation*}
\label{eq:SC_rules}
\setlength{\nulldelimiterspace}{0pt}
h(\bm{p}^h_{i\in\mathcal{I}}) =\left\{
\begin{IEEEeqnarraybox}[\relax][c]{l?sc}
0,  & if & 
\exists p_F:\sum_{i \in \mathcal{I}} p_{i,t}^h = p_{F}, \forall t \in \mathcal{T} . \\
+ \infty,  & \multicolumn{2}{l}{otherwise.}
\end{IEEEeqnarraybox}\right.
\end{equation*}
The third subproblem is an optimization per circle constraint~$s$, minimizing $g_s \left( \bm{z}^{g_s}_{i\in\mathcal{C}_s} \right), s = 1, \ldots n_{\mathcal{S}}$, with $g_s$ the indicator function of one constraint from (\ref{eq:opt_constr_circle}) - (\ref{eq:opt_constr_bool}) (i.e. with only the set $\mathcal{C}_s$ corresponding to circle constraint $s$).

To distribute the optimization problem, we use the alternating direction method of multipliers (ADMM) because of its superior convergence properties \cite{Boyd_ADMM} while being able to keep the cost functions local. 
However, as (\ref{eq:opt_problem}) is a mixed-integer problem and hence non-convex, ADMM nor other comparable distributed algorithms are guaranteed to converge to the global optimum~\cite{Eckstein1992}.
Nevertheless, we observe in Section~\ref{sec:casestudy} that the proposed distributed algorithm is able to converge to a suboptimal but feasible point in a finite number of iterations. 

To be able to distribute the problem using the ADMM methodology, each subproblem needs its own copy of the optimization variables $\bm{p}_i, \bm{z}_i$. 
Therefore, in the notation above and in what follows, we used the superscript $^f$ to denote the variables used in the local optimization of $f_i(\bm{p}_i^f, \bm{z_i}^f)$, the superscript $^h$ to denote the variables used in the optimization of $h(\bm{p}^h_{i\in\mathcal{I}})$ performed by the FSP and the superscript $^{g_s}$ to denote the variables used in the optimization of $g_s \left( \bm{z}^{g_s}_{i\in\mathcal{C}_s} \right)$ performed by the circle constraint agent $s$ managing the circle constraint $\mathcal{C}_s$.
This notation allows us to rewrite problem (\ref{eq:opt_problem}) as a consensus problem over the three subproblems:
\begin{mini!}
{\substack{(\bm{p}_i^f, \bm{z}_i^f, \bm{p}_i^{g_s},\\\bm{z}_i^{g_s}, \bm{p}_i^h)_{i \in \mathcal{I}}}}{ \sum_{i\in\mathcal{I}}f_i(\bm{p}_i^f, \bm{z}_i^f) +\sum_{\mathcal{C}_s \in \mathcal{S}} g_s \left( \bm{z}^{g_s}_{i\in\mathcal{C}_s} \right) + h \left( \bm{p}^h_{i\in\mathcal{I}} \right) }{\label{eq:opt_problem_dist}}{}\IEEEnonumber
\addConstraint{\bm{z}_i^f }{ =  \bm{z}_i^{g_s}, \quad}{  \forall i \in \mathcal{C}_s, \quad s=1,\ldots,n_{\mathcal{S}}   \label{eq:equal_constr_g} }
\addConstraint{\bm{p}_i^f }{ =  \bm{p}_i^h, \quad}{  \forall i \in \mathcal{I}.  \label{eq:equal_constr_h} }
\end{mini!}
Of every local binary vector $\bm{z}_i^f$, there is one copy per circle constraint $\bm{z}_i^{g_s}$ applicable to asset $i$ amongst which consensus has to be formed. The same holds for the local FCR capacity vector $\bm{p}_i^f$, of which there is a copy in the FSP objective $\bm{p}_i^h$.

To distribute (\ref{eq:opt_problem_dist}) using ADMM, one has to form the \emph{augmented Lagrangian} $\mathcal{L}_p$ of (\ref{eq:opt_problem_dist}):
\begin{IEEEeqnarray}{rCl}\label{eq:augm_lagr}
\mathcal{L}_p &=& \sum_{i\in\mathcal{I}} f_i(\bm{p}_i^f, \bm{z}_i^f) 
+\sum_{\mathcal{C}_s \in \mathcal{S}} g_s \left(\bm{z}^{g_s} _{i\in\mathcal{C}_s} \right) + h \left(\bm{p}^h_{i\in\mathcal{I}} \right) \nonumber \\
&&+ \sum_{\mathcal{C}_s \in \mathcal{S}} \sum_{i\in\mathcal{C}_s} (\rho_F/2)  \Vert \bm{z}_i^f -   \bm{z}_i^{g_s} + \bm{u}_i^{g_s}\Vert_2^2
  \\
&& +\sum_{i\in\mathcal{I}} (\rho_c/2) \Vert \bm{p}_i^f -   \bm{p}_i^{h} +\bm{u}_i^{h}\Vert_2^2. \nonumber
\end{IEEEeqnarray}
To keep the notation concise, we have used the scaled form of the augmented Lagrangian \cite{Boyd_ADMM}, with $\bm{u}_i^{g_s}$ and $\bm{u}_i^{h}$ the scaled dual variables corresponding to (\ref{eq:equal_constr_g}) and (\ref{eq:equal_constr_h}) respectively and $\rho_F,\rho_c >0$ the augmented Lagrangian parameters for the FSP constraint and the circle constraint, respectively. 

Note that this expression is slightly different from the traditional augmented Lagrangian, that only employs one value for rho: $\rho=\rho_F=\rho_c$. However, by allowing $\rho_F \neq \rho_c$, we are able to fine-tune the ADMM convergence as we are able to steer the convergence of the primal or dual residuals of the circle and the FSP constraints separately.

\begin{algorithm}
	\caption{Distributed ADMM optimization}
	\label{alg:ADMM}
	\begin{algorithmic}[1]
		\State $\bm{z}_i^{g_s}, \bm{u}_i^{g_s}\gets 0, \quad \forall i \in \mathcal{C}_s, s = 1,\ldots,n_\mathcal{S}.$ \label{alg:step1}
		\State$ \bm{p}_i^h, \bm{u}_i^h \gets 0, \quad  \forall i \in \mathcal{I}$.\label{alg:step2}
		\State $k\gets 0.$
		\While{ $\exists \mathcal{C}_s\in \mathcal{S}, t \in \mathcal{T} : \sum_{i\in\mathcal{C}_s} z^f_{i,t} > 10, z_{i,t}^f \in \{0, 1\} \label{alg_s:while1} $ \\ \hskip\algorithmicindent \hskip\algorithmicindent \textbf{and} $\Vert \sum_{i \in \mathcal{I}} ( \bm{p}^f_{i} - \bm{p}^h_{i} ) \Vert > \alpha \Vert \sum_{i \in \mathcal{I}} \bm{p}_i^f \Vert$} \label{alg_s:iter_k}
		\ForAll{$i \in \mathcal{I}$}  (in parallel) 
		\If{$(k \mod k_{IP}) = 0$}
		\State $\bm{p}_i^f, \bm{z}_i^f \gets \hat{\bm{p}}_i^f,  \hat{\bm{z}}_i^f, \ $ using (\ref{eq:local_opt}) with $z_{i,t}^f \in \{0, 1\}$.\label{alg_s:min_f_int}
		\Else
		\State $\bm{p}_i^f, \bm{z}_i^f \gets \hat{\bm{p}}_i^f,  \hat{\bm{z}}_i^f, \ $ using (\ref{eq:local_opt}). \label{alg_s:min_f}
		\EndIf
		\State \begin{varwidth}[t]{\linewidth}Send $\bm{z}_i^f$ to all circle constraint agents $s :  \mathcal{C}_s \in \mathcal{S}_i$ \\ \hskip\algorithmicindent \hskip\algorithmicindent  and $\bm{p}_i^f$ to the FSP. \label{alg_s:sent_local}
		\end{varwidth}
		\EndFor 
		\ForAll{$\mathcal{C}_s \in \mathcal{S}$}  (in parallel)
		\State \raggedright $\bm{z}_i^{g_s}\gets \hat{\bm{z}}_i^{g_s}, \quad \forall i \in \mathcal{C}_s, \quad$ using (\ref{eq:local_circle_opt}). \label{alg_s:circle_opt} 
		\State $\bm{u}_i^{g_s} \gets \bm{u}_i^{g_s} + \bm{p}_i^f - \bm{z}_i^{g_s}, \quad \forall i \in \mathcal{C}_s.$ \label{alg_s:update_dual_circle}
		\State Send $\bm{z}_i^{g_s}, \bm{u}_i^{g_s}$ to all assets $i \in \mathcal{C}_s$. \label{alg_s:sent_circle}
		\EndFor
		\State $\bm{p}_i^{h}\gets \hat{\bm{p}}_i^{h}, \quad \forall i \in \mathcal{I}, \quad$ using (\ref{eq:FSP_opt}). \label{alg_s:fsp_opt} 
		\State $\bm{u}_i^{h} \gets \bm{u}_i^{h} + \bm{p}_i^f - \bm{p}_i^{h}, \quad \forall i \in \mathcal{I}$. \label{alg_s:update_dual_fsp}
		\State Send $\bm{p}_i^{h}, \bm{u}_i^{h}$ to all assets $i \in \mathcal{I}$. \label{alg_s:fsp_sent}	
		\State $k\gets k + 1 $.
		\EndWhile
	\end{algorithmic}
\end{algorithm}

The proposed distributed optimization algorithm, shown in full in Algorithm \ref{alg:ADMM}, consists of the alternating partial minimization of the augmented Lagrangian (\ref{eq:augm_lagr}). First, an optimization is performed by the local agents minimizing $f_i$, followed by the circle constraint agents minimizing $g_s$ and the FSP minimizing $h$. 
We will now discuss the three subproblems in detail.

\subsubsection{Local Optimization}
In the first step of each iteration~$k$, (step \ref{alg_s:min_f_int} or \ref{alg_s:min_f} in Algorithm \ref{alg:ADMM}), the augmented Lagragian is minimized for each asset $i \in \mathcal{I}$ over the FCR capacity $\bm{p}_i^f$ and binary $\bm{z}_i^f$ variables, while keeping the other variables constant. 
The local optimization, to be executed in parallel at each asset $i$, can then be formulated as:
\begin{argmini!}
{\bm{p}_i^f, \bm{z}_i^f }{ c_{i}(\bm{p}_{i}^f)  +  \sum_{\mathcal{C}_s \in \mathcal{S}_i} \frac{\rho_F}{2}  \Vert \bm{z}_i^f - \bm{z}_i^{g_s} + \bm{u}_i^{g_s}\Vert_2^2  }{\label{eq:local_opt}}{\hat{\bm{p}}_i^f,  \hat{\bm{z}}_i^f = }\IEEEnonumber
\breakObjective{  + \frac{\rho_c}{2} \Vert \bm{p}_i^f - \bm{p}_i^{h} +\bm{u}_i^{h}\Vert_2^2 - c_F\sum_{t\in\mathcal{T}}  p_{i,t}^f}
\addConstraint{0\leq}{ {p}_{i,t}^f \leq 5 z_{i,t}^f , \quad\forall t \in \mathcal{T} } 
\addConstraint{\bm{z}_i^f}{ \in [0, 1]^{n_\mathcal{T}}. \label{eq:localopt_int}}
\end{argmini!}
This optimization problem incorporates the local optimization $f_i$, given by the part of the objective (\ref{eq:opt_problem_obj}) applicable to asset $i$ constrained to~(\ref{eq:opt_problem_constr}), together with the quadratic penalty terms of the augmented Lagragian (\ref{eq:augm_lagr}) which are applicable to asset~$i$. As this optimization only considers the variables of asset $i$ and does not need variables of other assets, it succeeds in keeping all information about cost and local constraints private. 


In (\ref{eq:local_opt}), we have relaxed the binary constraint (\ref{eq:opt_constr_bool}) to be continuous, thereby avoiding oscillatory behaviour between this binary variable $z_{i,t}^f$ and the binary variables from~(\ref{eq:local_circle_opt}), $z_{i,t}^{g_s}$. However, with $z_{i,t}^f$ continuous, the local asset $i$ does not know if it actually can provide any FCR capacity. Therefore, every $k_{IP}$ iterations, problem (\ref{eq:local_opt}) is solved with $z_{i,t}^{g_s}$ constrained to an integer variable (step \ref{alg_s:min_f_int}).



 

When asset $i$ finishes its local optimization, in step \ref{alg_s:sent_local}, the FCR capacity variables $\hat{\bm{p}}_i^f$ are sent to the FSP for the optimization of $h$. The binary variables $\hat{\bm{z}}_i^f$ are sent to the circle constraint agents $s : \mathcal{C}_s \in \mathcal{S}_i$ of all circle constraints of which asset $i$ is part of.

As shown in steps \ref{alg:step1}-\ref{alg:step2} of algorithm \ref{alg:ADMM}, we initialize the algorithm by setting $\bm{z}_i^{g_s} = \bm{p}_i^{h} = 0$.
However, with $\rho_F, \rho_c > 0$ this results in a couple of initial iterations needed to raise $\bm{p}_i^f$ from 0 towards their economical value.
These initial iterations can be avoided by setting $\rho_F = \rho_c = 0$ in the first iteration, thereby \emph{warm-starting} the algorithm, as this means solving the optimization problem (\ref{eq:opt_problem}) without applying constraints (\ref{eq:opt_constr_global}), (\ref{eq:opt_constr_circle}). 
This immediately results in using all assets at maximum FCR capacity ${p}_{i,t}^f = 5$ if this is economically interesting, i.e. if $c_F \sum_{t\in\mathcal{T}}  p_{i,t} > c_{i}(\bm{p}_i)$.

\subsubsection{Circle Constraint Optimization}
Having received all locally optimized binary variables $\bm{z}_i^f, i \in \mathcal{C}_s$, in step \ref{alg_s:circle_opt} each circle constraint agent $s$ optimizes (\ref{eq:augm_lagr}) over its copy of these variables $\bm{z}_{i \in \mathcal{C}_s}^{g_s}, $ to respect the circle constraints (\ref{eq:opt_constr_circle})-(\ref{eq:opt_constr_bool}), corresponding to the partial optimization of the augmented Lagrangian (\ref{eq:augm_lagr}). 
This optimization is also separable and can thus be executed for every circle constraint $s$ in parallel.
The optimization problem of one such circle constraint agent $s$ results in a quadratic mixed-integer program:
\begin{argmini}
{\bm{z}_{i\in\mathcal{C}_s}^{g_s}}{\sum_{i\in\mathcal{C}_s} (\rho_c/2)  \Vert \bm{z}_i^f -   \bm{z}_i^{g_s} + \bm{u}_i^{g_s}\Vert_2^2}{\label{eq:local_circle_opt}}{\hat{\bm{z}}_{i\in\mathcal{C}_s}^{g_s} = }
\addConstraint{\sum_{i\in\mathcal{C}_s} z_{i,t}^{g_s}}{\leq 10, \quad}{ \forall t \in \mathcal{T}}
\addConstraint{\bm{z}_i^{g_s}}{ \in \{0, 1\}^{n_\mathcal{T}},  \quad}{ \forall i \in \mathcal{C}_s.} 
\end{argmini}
Note that this problem is actually the Euclidean projection of $(\bm{z}_i^f + \bm{u}_i^{g_s})$ on the feasible set defined by constraints (\ref{eq:opt_constr_circle})-(\ref{eq:opt_constr_bool}).

As (\ref{eq:local_circle_opt}) is a quadratic mixed-integer problem, it can be hard to solve. 
However, as there is no coupling in time, the problem can be separated into $n_\mathcal{T}$ distinct subproblems, each containing only one binary variable $z_{i,t}^t$, which can be solved very efficiently.

Having obtained the optimal $\hat{\bm{z}}_i^{g_s}$, the scaled dual variables of each point of each circle constraint $\bm{u}_i^{g_s}$ are updated in step~\ref{alg_s:update_dual_circle}. Both updates are sent back to the corresponding assets $i$ in step~\ref{alg_s:sent_circle} for use in the next iteration $k+1$.

\subsubsection{FSP Optimization}
The FSP tries to obtain a constant FCR capacity $p_F$ from all assets over the bidding period $n_{\mathcal{T}}$.
At iteration $k$, the FSP gathers all locally optimized variables $\bm{p}_i^f$ from all assets $i\in \mathcal{I}$ and minimizes the augmented Lagrangian (\ref{eq:augm_lagr}) over its copy of the local FCR capacity variables~$\bm{p}_i^h, i \in \mathcal{I}$:
\begin{argmini}
{p_F, \bm{p}^{h}_{i\in \mathcal{I}}}{ \sum_{i \in \mathcal{I}}(\rho_F/2) \Vert \bm{p}_i^f -   \bm{p}_i^{h} +\bm{u}_i^{h}\Vert_2^2 
}{\label{eq:FSP_opt}}{\hat{\bm{p}}^{h}_{i\in \mathcal{I}} = }
\addConstraint{\sum_{i\in\mathcal{I}} p_{i,t}^h}{= p_F,\quad   }{ \forall t \in \mathcal{T}.}
\end{argmini}
This problem is independent of (\ref{eq:local_circle_opt}), and can thus be executed in parallel to (\ref{eq:local_circle_opt}).

As with the circle constraints, when having obtained the optimal $\bm{p}_i^{h}$, the scaled dual variables $\bm{u}_i^{h}$ are updated in step~\ref{alg_s:update_dual_fsp}. Finally, in step~\ref{alg_s:fsp_sent}, both $\bm{p}_i^{h}$ and $\bm{u}_i^{h}$ are sent back to the corresponding assets $i$.


When asset $i$ has received both $\bm{p}_i^{h}, \bm{u}_i^{h}$ from the FSP and  $\bm{z}_i^{g_s}, \bm{u}_i^{g_s}$ from all circle constraint agents $s: \mathcal{C}_s \in \mathcal{S}_i$, it can start its next iteration $k+1$ by solving (\ref{eq:local_opt}) with the updated parameters $\bm{p}_i^{h}, \bm{u}_i^{h}, \bm{z}_i^{g_s}, \bm{u}_i^{g_s}$, until convergence is reached. 

The algorithm converges when the local variables $\bm{p}_i^f, \bm{z}_i^f$ reach a feasible operating point, as denoted by the while-conditions in steps~\ref{alg_s:while1} and \ref{alg_s:iter_k}. 
The condition in step~\ref{alg_s:while1} evaluates the circle constraints (\ref{eq:opt_constr_circle}) with $z_{i,t}^f$ binary, and can thus only be checked every $k_{IP}$ iterations. The condition in step~\ref{alg_s:iter_k} evaluates if the total local FCR capacity $\sum_{i \in \mathcal{I}} p_{i,t}^f$ is close enough to the aggregated FCR capacity $p_F = \sum_{i \in \mathcal{I}} p_{i,t}^h$, with $\alpha$ a parameter denoting the relative error between the norm of the difference $\sum_{i \in \mathcal{I}} ( \bm{p}^f_{i} - \bm{p}^h_{i} )$ over every time step $t$, relative to the norm of the total local FCR capacity $\sum_{i \in \mathcal{I}} \bm{p}_{i}^f$.


\section{Case Study: Distributed Assets in Breda}\label{sec:casestudy}
In this section, we evaluate the impact of the circle constraints on the portfolio capacity an FSP can provide with low-voltage connected, distributed assets providing FCR and the performance of the proposed distributed ADMM algorithm. 

We present a case study using actual data of the low-voltage connection points in the city of Breda, a middle-sized municipality in the South of the Netherlands with 183\,765 inhabitants~\cite{bevolkingsregister} and 86\,868 LV connection points. The geographic location of the connection points in the municipality is provided by Enexis, the local distribution grid operator, with an accuracy up to \SI{1}{\metre} and can be freely accessed online~\cite{Enexis_data}.
We used the municipality of Breda as a case study rather than a Belgian municipality as accurate data on the location of LV connection points was not available for a Belgian municipality.

\subsection{Impact of Circle Constraints on Usable FCR Capacity}

\begin{figure}
	\centering
	\includegraphics[width=\columnwidth]{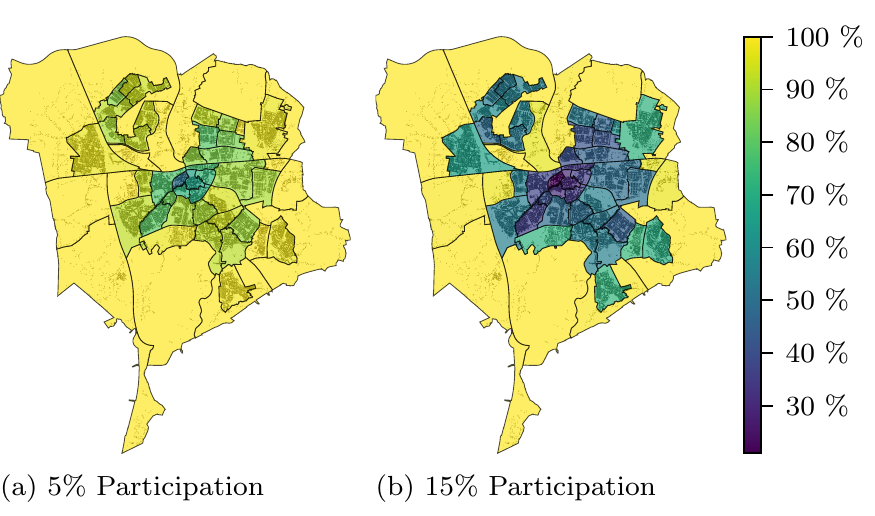}
	\vspace*{-0.7cm}
	\caption{Usable proportion of total available LV grid connected FCR capacity in different neighbourhoods in the city of Breda, taking into account the circle constraints imposed by Synergrid, in case 5\% and 15\% of all LV connection points are able to participate in FCR.} 
	\label{fig:choropleth}
		\vspace*{-0.3cm}
\end{figure}

\begin{figure}
	\centering
	\subfloat{\centering \hspace{0.6cm}
		\includegraphics[width=0.90\columnwidth]{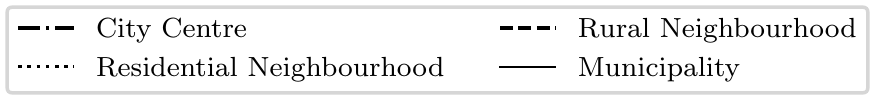}
	}\vspace*{-0.3cm}
	\setcounter{subfigure}{0}
	\subfloat[Usable proportion of total available FCR capacity. \label{fig:usable_prop}]{%
		\includegraphics[width=0.475\columnwidth]{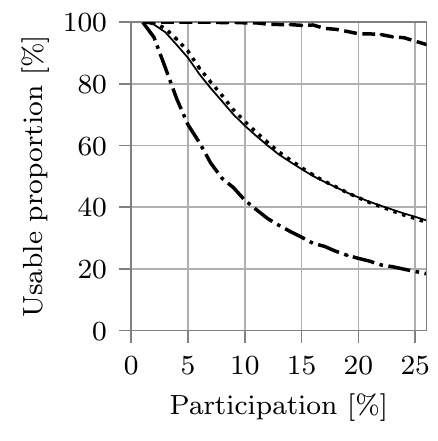}
	}\hfill
	\subfloat[Total FCR capacity.\label{fig:cap_neigh}]{%
		\centering
		\includegraphics[width=0.475\columnwidth]{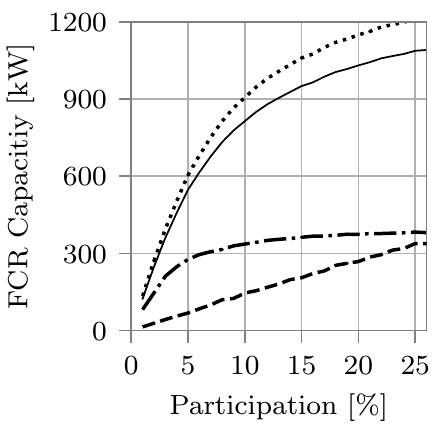}
	}
	\caption{(a) Usable proportion of total available FCR capacity and (b) total FCR capacity in function of the participation rate in the averaged city centre, residential, rural neighbourhood and of the entire municipality.}
	\label{fig:cap_part}
	\vspace*{-0.3cm}
\end{figure}

The circle constraints imposed by Synergrid will reduce the amount of FCR capacity that can be used by the FSP.
Figure~\ref{fig:choropleth} shows the maximal proportion of the total available FCR capacity that can be used in an FSP pool per neighbourhood in Breda, taking into account the circle constraints as imposed by Synergrid. 
Figure \ref{fig:choropleth}a gives the results for a participation rate of \SI{5}{\%}, meaning that \SI{5}{\%} of all available LV connection points are able to provide FCR capacity, while Figure \ref{fig:choropleth}b gives the results for a participation rate of \SI{15}{\%}.
In the figure, a neighbourhood with a value of \SI{100}{\%} indicates that all available FCR capacity, and thus all participating assets, can be used in the FSP pool, while a value of e.g. \SI{70}{\%} indicates that only up to \SI{70}{\%} of the available FCR capacity or participating assets can be used for FCR, in order to comply with the circle constraints imposed by Synergrid.
We select the location of the participating assets randomly from the available connection points according to a uniform distribution, so that the total number of participating assets equals the participation rate. 

For this calculation, we assumed that every participating connection point is able to provide at least \SI{5}{kW} of FCR capacity and selected the participating connection point randomly from all LV connection points in Breda. 
We calculated the maximum amount of FCR capacity in an FSP pool can by solving (\ref{eq:opt_problem}) with $ \forall i: c_{i}(\bm{p}_i) = 0$ and $c_F=1$, for which we used Gurobi~\cite{gurobi}.
To eliminate the effect of randomness in the location of the assets, Figure \ref{fig:choropleth} shows the Monte Carlo average of ten such simulations.

From Figure \ref{fig:choropleth}, it is clear that there exists a large spatial difference in reduction of FCR capacity: at \SI{5}{\%} participation, mainly in the city centre there is already a substantial amount of reduction of the usable FCR capacity. At \SI{15}{\%} participation, the reduction of the usable FCR in the city centre becomes very strong and also in more suburban neighbourhoods there is a considerable amount of reduction observable. On the other hand, in the more rural, outer neighbourhoods, up to \SI{100}{\%} of available FCR capacity can still be used.

Figure \ref{fig:usable_prop} quantifies this spacial difference, showing the usable proportion of available FCR capacity in function of the degree of participation for the three types of neighbourhoods identified previously: rural, residential (suburban) and the city centre. The neighbourhoods in the city centre show the biggest decrease between \SI{2}{\%} and \SI{10}{\%} participation, while rural neighbourhoods only start to decrease their usable proportion after \SI{15}{\%} participation, at a much slower rate.

However, this slower reduction in rural neighbourhoods does not automatically translate into a higher total FCR capacity, as the number of connection points in a rural neighbourhood is much lower than in a residential neighbourhood. This can be seen in Figure \ref{fig:cap_neigh}, which shows the total FCR capacity of each type of neighbourhood. While FCR capacity increases linearly with the participation rate in rural neighbourhoods, the capacity is still much lower than in a residential neighbourhood, as there are many more connection points. 
In the city centre, the increase in total FCR capacity levels off at around \SI{5}-\SI{10}{\%} participation, indicating that any additional FCR assets deployed in the city centre beyond this level of participation will not increase the total FCR capacity of the pool of the FSP.

\subsection{Performance of the Distributed Optimization Algorithm}
This section discusses the performance of the proposed distributed optimization algorithm. 
To limit the simulation time, 
we limit ourselves to simulating only the \emph{Zandberg} neighbourhood in Breda (shown in Figure~\ref{fig:circles}), a typical residential neighbourhood.

We simulate $n_\mathcal{T} = 24$ time steps, each corresponding to one hour of a day. 
To keep the optimization problems efficiently solvable, we assume the local cost of providing FCR capacity $c_{i}(\bm{p}_i)$ is a linear function of the frequency control capacity: $c_{i}(\bm{p}_i) = \tilde{\bm{c}}_i^\intercal \bm{p}_i$, with $\tilde{\bm{c}}_i = (\tilde{c}_{i,0},\ldots, \tilde{c}_{i,n_\mathcal{T}})^\intercal$.
As the goal is to obtain an idea of the performance of the algorithm, rather than calculating the actual monetary value of the objective, we choose $\tilde{c}_{i,t}$ from a uniform distribution between 0 and 1 for each time step~$t$ and asset~$i$.
The price of FCR capacity is chosen to be $c_{FCR} = 0.8$, as primary frequency control is usually one of the most valuable services for flexibility~\cite{Oudalov2006}. Choosing $c_{FCR} = 0.8$, means that in \SI{20}{\%} of the cases the local cost for FCR capacity will still be higher than what one can gain from FCR capacity during that time step. 
To compare the performance of the distributed optimization algorithm, we also solve problem (\ref{eq:opt_problem}) in a centralized fashion towards a global optimal point, using the Gurobi solver\cite{gurobi}.

Finally, in Algorithm~\ref{alg:ADMM}, we set $k_{IP}=10$ and $\alpha = 0.005$.


\subsubsection{Convergence}
\begin{figure}
	\centering
	\subfloat[Objective value (\ref{eq:opt_problem_obj}) of the distributed and centralized optimization. \label{fig:obj}]{%
		\includegraphics[width=0.475\columnwidth]{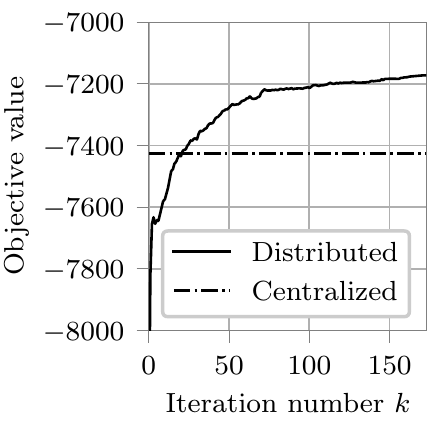}
	}\hfill
	\subfloat[Euclidean norms of the primal residuals of  circle constraints (\ref{eq:opt_constr_circle}) and the FSP constraint~(\ref{eq:opt_constr_global}). \label{fig:primal_res}]{%
		\centering
		\includegraphics[width=0.475\columnwidth]{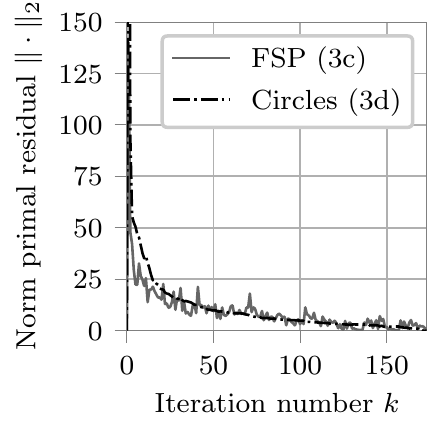}
	}
	\caption{Evolution of the objective value (a) and primal residuals of the distributed constraints (b) using the proposed distributed optimization, in function of the iteration number $k$.}
	\label{fig:conv}
		\vspace*{-0.3cm}
\end{figure}

Figure \ref{fig:obj} the evolution of the objective value when simulating the distributed algorithm with $\rho_F=0.25$ and $\rho_c=0.3$, for \SI{10}{\%} participation in function of the iteration number $k$. The figure also shows the global optimum of the centralized solver.
As expected, rather than to the global optimum, the distributed optimization converges to a suboptimal solution which is \SI{3.4}{\%} from the global optimum. 

Figure~\ref{fig:primal_res} shows the evolution of the Euclidean norm of the primal residuals of constraints (\ref{eq:opt_constr_circle}) and (\ref{eq:opt_constr_global}), the two constraints that have been distributed. One can see that both norms decrease rapidly in the beginning, but convergence slows down when the algorithm advances. For this scenario, the algorithm converges to a feasible point in 173 iterations. 

Comparing Figure \ref{fig:obj} with Figure \ref{fig:primal_res}, one can observe that the low objective value in the beginning is possible due to the high value of the primal residual of constraints (\ref{eq:opt_constr_global}) and  (\ref{eq:opt_constr_circle}). When the distributed optimization advances, the solution is forced towards a more feasible solution, thereby increasing the objective value and decreasing the primal residuals. The values of $\rho_c$ and $\rho_F$ determine how fast the primal residuals are forced towards zero.





\subsubsection{Impact of $\rho_c$}
\begin{figure}
	\centering
	\includegraphics[width=0.9\columnwidth]{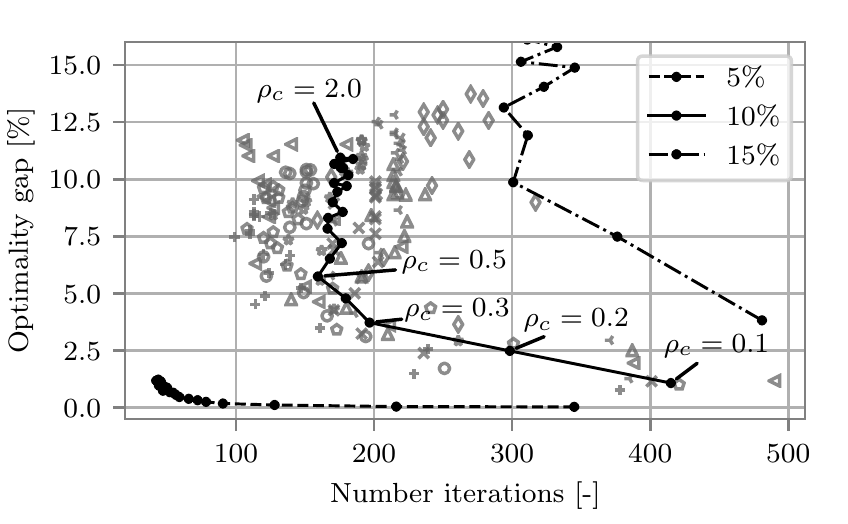}
	\vspace*{-0.3cm}
	\caption{Optimality gap versus number of iterations until convergence for various values of $\rho_c$, at \SI{10}{\%} participation. Each marker type represents a different randomly drawn scenario, ran for $\rho_c$ ranging between 0.1 and 2.0 and $\rho_F = 0.25$. The black line is the average of all scenarios, with each black dot corresponding to one value of $\rho_c$. The dashed lines are the averages at \SI{5}{\%} and \SI{15}{\%} participation.}
	\label{fig:opt_gaps}
	\vspace*{-0.3cm}
\end{figure}
Figure \ref{fig:opt_gaps} shows the relative optimality gap of the distributed optimization algorithm versus the number of iterations until convergence towards a feasible point, for values of $\rho_c$ ranging between 0.1 and 2.0 and $\rho_F=0.25$. 
The relative optimality gap is defined as the difference between the objective value of the converged distributed optimization algorithm and the global optimum found by the centralized solver, divided by this global optimum.
To account for the randomness in the location of the assets and the local costs $c_{i,t} $, we ran the distributed optimization algorithm for eight randomly drawn scenarios.
Each marker type in Figure \ref{fig:opt_gaps} represents one scenario with \SI{10}{\%} participation with the continuous black line the average of the eight scenarios. As can be seen, there is a clear trade-off between the optimality gap and the number of iterations needed for convergence, which can be tuned by varying the parameter $\rho_c$. 

Increasing $\rho_c$ results in quicker convergence towards a feasible point as more weight is put on the penalty terms in the augmented Lagrangian (\ref{eq:augm_lagr}), but the algorithm also gets stuck quicker on a suboptimal integer solution. A smaller $\rho_c$ on the other hand encourages exploration of other integer solutions, but also requires more iterations to reach a feasible point.

As $\rho_c$ only impacts convergence of the circle constraints (\ref{eq:opt_constr_circle}), an analogue reasoning can be followed for $\rho_F$ and the convergence of the FSP constraint (\ref{eq:opt_constr_global}). However, as the circle constraints are much more restricting, we found that the impact of $\rho_F$ is much less than that of $\rho_c$.

Figure \ref{fig:opt_gaps} also shows the averages of eight randomly drawn scenarios with \SI{5}{\%} participation and five scenarios with \SI{15}{\%} participation. These show a clear trend: as the number of participating assets decreases, both the optimality gap and the number of iterations needed for convergence decrease as well, and vice versa. However, the trade-off between iterations and optimality gap remains visible for all participation degrees.

Although the optimality gap increases with a higher participation, this will not cancel out the additional revenues resulting from a larger aggregate FCR capacity $p_F$ that comes with a higher participation, as indicated by Table \ref{tbl:objs}. 
The table shows the global optimum found with a centralized solver and the objective value of the distributed algorithm with $\rho_c=0.3$ after convergence, averaged over all simulated scenarios. As one can see, although the optimality gap increases with increasing participation, the objective value of the distributed optimization still decreases (which means an increase in revenues) due to a larger amount of FCR capacity that can be valorized.

\begin{table}[h]
	\renewcommand{\arraystretch}{1.3}
	\centering
	\caption{Average Optimality Gap and Objective Value (\ref{eq:opt_problem_obj}), Solved to a Global Optimum with a Centralized Solver and with the Distributed Optimization Algorithm with $\rho_c=0.3$. }
	\label{tbl:objs}
	\begin{tabular}{l|rrr}
		\toprule
		& \multicolumn{3}{c}{Participation [\%]}\\ 
		& \multicolumn{1}{c}{5}   & \multicolumn{1}{c}{10}   & \multicolumn{1}{c}{15}   \\ \midrule
		Objective (\ref{eq:opt_problem_obj}) global optimum  & -4201 & -7451 & -9464\\
		Objective (\ref{eq:opt_problem_obj}) distributed optimization & -4194 & -7175 & -8532 \\
		Optimality gap distributed optimization & 0.18\% & 3.2\% & 9.9\% \\
		\bottomrule
	\end{tabular}
\end{table}

\begin{figure}
	\centering
	\includegraphics[width=0.95\columnwidth]{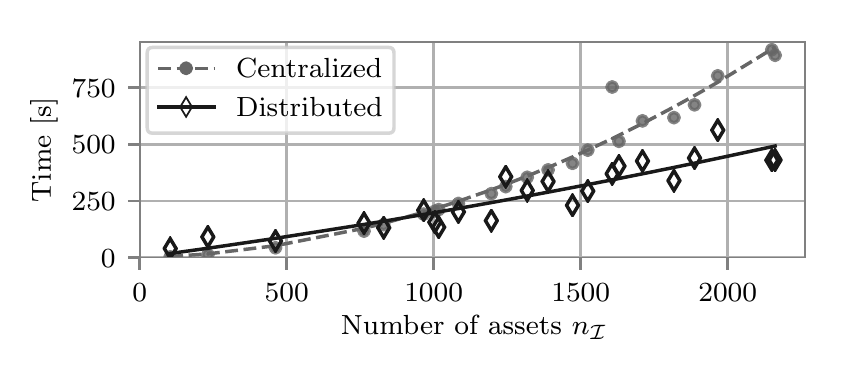}
	\vspace*{-0.3cm}
	\caption{Optimization time in function when solving the problem centralized, using cvxpy~\cite{cvxpy} and Gurobi~\cite{gurobi} and with the distributed optimization algorithm with $\rho_c=0.3$ and a participation of \SI{10}{\%}.}
	\label{fig:time}
	\vspace*{-0.3cm}
\end{figure}

Finally, Figure~\ref{fig:time} compares the execution time of the distributed algorithm with a centralized solver, in function of the number of assets $n_\mathcal{I}$ in the simulation. The results are obtained by including an increasing number of neighbourhoods, each with a participation of \SI{10}{\%}, into the simulation.
The time needed for the centralized solver increases rapidly with the number of nodes, while the time for the distributed solver increases at a slower rate, owing to the parallelization of the local optimization (\ref{eq:local_opt}) and of the circle constraint optimization~(\ref{eq:local_circle_opt}). The figure clearly shows that, as from about 1000 assets, the distributed optimization becomes faster than a centralized solver.

In \cite{FEIZOLLAHI201597}, another heuristic is proposed for ADMM applied to a mixed-integer unit commitment problem: first solving the continuous relaxed problem, 
and then switching to the integer problem. 
However, solving the continuous relaxed problem 
already consumes a lot of iterations, and we found that a same optimality gap can be achieved with fewer iterations by choosing an appropriate value of $\rho_c$.


\section{Conclusion}\label{sec:conclusion}

In this paper, we presented the problem of an FSP operating a pool of low-voltage connected flexibility resources to optimally provide FCR, while being compliant with the 2018 Belgian regulatory constraints at the distribution grid level.

We showed that these new regulatory constraints have a considerable impact on the total FCR capacity that can be monetized, however the impact varies strongly between neighbourhoods with different population densities.

We elaborated the mixed-integer problem of an FSP when operating such a pool of low-voltage grid connected assets and have proposed a distributed optimization algorithm that solves the problem in a tractable way while maintaining confidentiality of costs and the constraints of the local participants. 
A performance assessment of the distributed optimization shows a trade-off between the number of iterations needed for convergence towards a feasible point and the optimality gap. We showed that the distributed optimization converges quicker than a centralized solver when the number of participating assets increases above 1000.

Future work includes comparison of the proposed distributed ADMM with other distributed algorithms, such as the dual ascent method or column generation and an assessment on how the optimality gap could be decreased. 
Finally, it would be interesting to 
assess to what extent the new regulations are able to mitigate problems in the low-voltage distribution grid.






%
\bibliographystyle{IEEEtran}
\bibliography{IEEEabrv,bibl_R1_LS}{}

\end{document}